\documentclass[fleqn,12pt]{wlscirep}
\usepackage[utf8]{inputenc}
\usepackage[T1]{fontenc}
\usepackage{cancel}
\usepackage[normalem]{ulem} 
\usepackage{soul} 
\title{Coherent electron displacement for quantum information processing using attosecond single cycle pulses}

\author[1,*]{Hicham Agueny}

\affil[1]{Department of Physics and Technology, Allegt. 55,
University of Bergen, N-5007 Bergen, Norway}

\affil[*]{hicham.agueny@uib.no}

\begin{abstract}
Coherent electron displacement is a conventional strategy for processing quantum information, as it enables to interconnect distinct sites in a network of atoms. The efficiency of the processing relies on the precise control of the mechanism, which has yet to be established. Here, we theoretically demonstrate a new route to drive the electron displacement on a timescale faster than that of the dynamical distortion of the electron wavepacket by utilizing attosecond single-cycle pulses. The characteristic feature of these pulses relies on a vast momentum transfer to an electron, leading to its displacement following a unidirectional path. The scenario is illustrated by revealing the spatiotemporal nature of the displaced wavepacket encoding a quantum superposition state. We map out the associated phase information and retrieve it over long distances from the origin. Moreover, we show 
that a sequence of such pulses applied to a chain of ions enables attosecond control of the directionality of the coherent motion of the electron wavepacket back and forth between the neighbouring sites. An extension to a two-electron spin state demonstrates the versatility of the use of these pulses. Our findings establish a promising route for advanced control of quantum states using attosecond single-cycle pulses, which pave the way towards ultrafast processing of quantum information as well as imaging.
\end{abstract}
\begin{document}

\flushbottom
\maketitle

\thispagestyle{empty}

\section*{Introduction}

Nowadays, single atoms become the building blocks for the emerging quantum technology, in particular, for developing atom-based electronic devices for processing signal and information~\cite{Wyrick2019}.
As individual electrons carry information encoded in their wavefunction~\cite{McNeil2011,Bertrand2016,Bauerle2018}, they are thus exploited in a variety of applications in quantum electronics notably for processing quantum information~\cite{McNeil2011}. It thus becomes apparent that a mechanism capable of transferring electrons between different sites in a network of atoms is required~\cite{McNeil2011}. In this context, electron displacement has been suggested as a conventional strategy for processing quantum information efficiently~\cite{McNeil2011,Flentje2017}. The efficiency relies on the precise control of the transferring electrons over long distances~\cite{Flentje2017}. In general, the ultimate coherent control requires the study of the strong light-matter interaction occurring on a short timescale against the decoherence~\cite{Katsuki2013}. Driving the electron displacement on an ultrashort timescale is, therefore, the key for achieving ultrafast processing of quantum information with high-fidelity. 

Here we present a scheme to drive the coherent electron displacement on an unprecedented timescale by utilizing an intense single-cycle pulse. This class of pulses has emerged due to a need for a precise control of the coherent electron motion~\cite{Krauss2010,Tibai2014,Tanaka2015,Balciunas2015,Liang2017,Mak2018,Nie2018,Hwang2019}. The characteristic feature of these pulses was linked to a momentum kick that a bound electron receives from the single-cycle field~\cite{Jones2014,Robicheaux2014}. As a result, the electron is ionized mostly, but not completely, in a single-direction~\cite{Agueny2016,Misha2017a}, and the underlying mechanism is identified as a non-zero displacement ionization~\cite{Robicheaux2014}. This ionization mechanism defines another important regime beyond the tunneling and multiphoton ionization, which are widely explored in strong-field physics. The particularity of these pulses has been already discussed a decade ago~\cite{Briggs2008}: using the first order Magnus expansion, it was suggested that a strong single-cycle field may effectively displace the electron wavepacket far away without changing its wavefunction, unlike in previous studies~\cite{Robicheaux2014,Agueny2016,Misha2017a}, thus leading to a formation of an "atom without nucleus"~\cite{Briggs2008}. This prediction, although is based on a perturbative approach, it gets support from recent preliminary non-perturbative calculations based on a one-dimensional model and deeply analysed in the Kramers-Henneberger frame~\cite{Micha2017b}. 

Despite the unique feature of these pulses for inducing coherent displacement of the electron, and the rapid progress of their generation, theoretical considerations of the mechanism, so far, are scarcely and were previously limited to simplified models~\cite{Briggs2008,Micha2017b}. Specifically, there has been no theoretical works on resolving in time domain the electron displacement effect. It is thus timely to guide experiments for exploring new routes when attosecond single-cycle pulses may promote cutting-edge applications in strong-field and attosecond physics~\cite{Nie2018}. 

We demonstrate the time-resolved and control of the spatial coherent electron displacement on the attosecond (as) up to femtosecond (fs) timescales, which has not yet established; thus adding new insights to the general field of strong-field and attosecond physics. This is achieved on the basis of ab-initio calculations for quantum systems having one and two electrons. The scenario is demonstrated by illustrating a scheme in which a quantum superposition state, which is considered to be prepared by a pump laser pulse in a coherent mixture of two Rydberg states, can be transported far away from the origin following a unidirectional path and without loss of coherence. The loss of coherence is related here to the vanishing of the overlap between the two Rydberg states due to a change of the phase coupling between these two states caused by the probing pulse. With the help of a simple physical model, we are able to map out the associated phase information encoded in the electron wavefunction, and retrieve it over long distances from the origin. An extension to trapped ions, in which atom-based devices rely on, shows that a sequence of such pulses enables an efficient attosecond transport of the electron wavepacket back and forth between the neighbouring sites, thus offering opportunities for ultrafast electronic communication between quantum dot nanostructures. The scenario demonstrated here is not limited to a single-electron system, but is further shown to be applicable to a two-electron spin dynamics, thus demonstrating the versatility of the use of attosecond single-cycle pulses. 

\begin{figure}
\centering
\includegraphics[width=9cm,height=8cm]{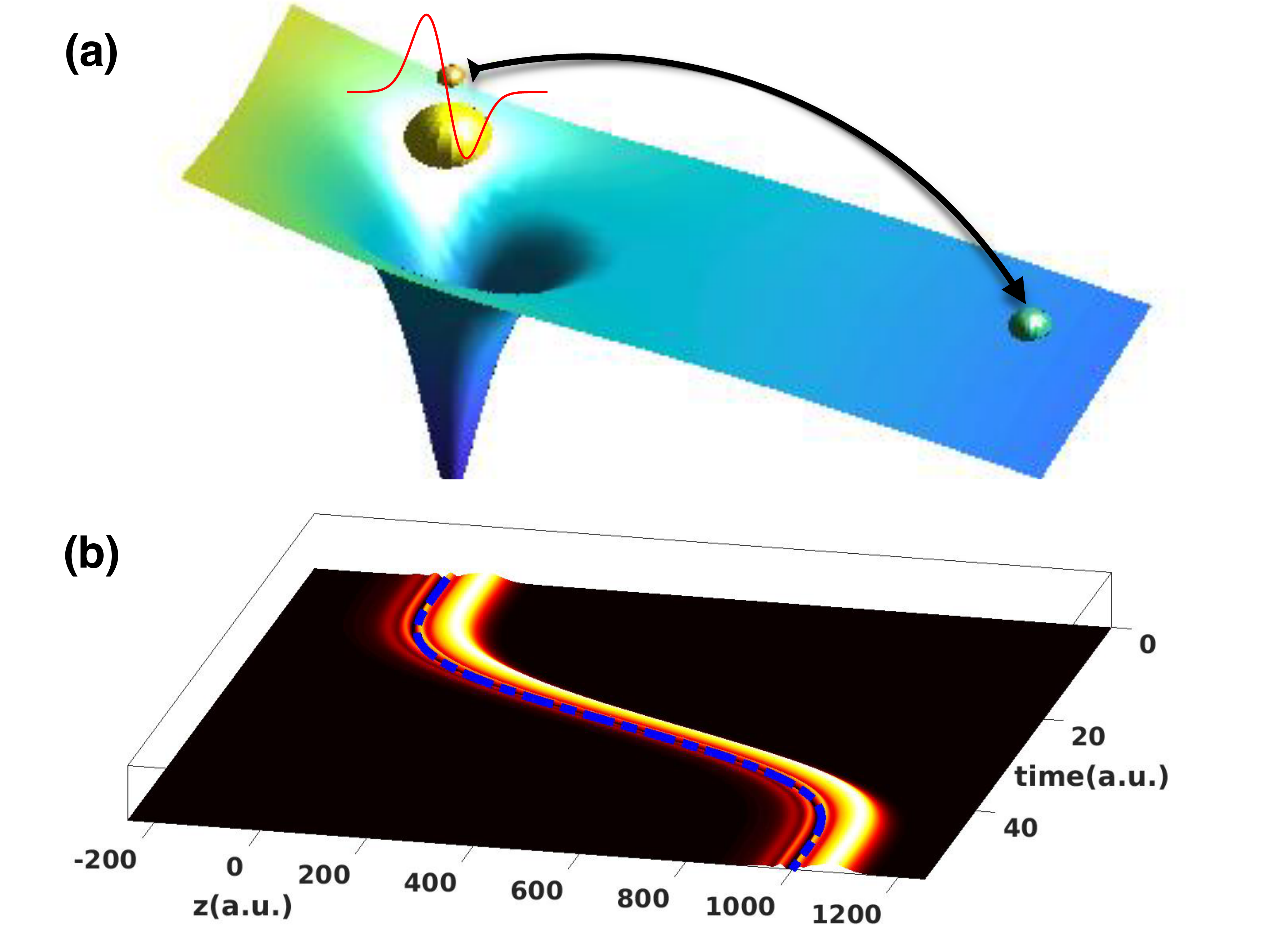}
\caption{\label{fig1} Schematic description of the displacement mechanism from an atomic target. (a) Illustration of the mechanism: the electron receives a momentum kick from a single-cycle pulse (red curve) and is displaced far away from the nucleus. (b) The temporal evolution of the electron density integrated over the $\rho$-direction for an initial phase of $\phi(t_i)=0$ and for the polar angle $\theta_R=\pi/4$. The displacement vector $\alpha(t)$ is shown with dashed blue line. The peak intensity is fixed at 4.26 10$^{18}$ W/cm$^2$ and the angular frequency is 0.057 a.u..}
\end{figure}

\section*{Results}
\subsection*{Coherent mixture of two states}
To demonstrate the feasibility of a coherent transfer of a superposition state, we consider an intense single-cycle pulse to probe the mixture of the involved states. The mechanism here is based on the momentum transfer between the single-cycle field and the bound electron. The latter receives a momentum kick from the field and is left far away from its origin in a well-defined region in position space, as illustrated in Fig.~\ref{fig1}(a). In the scheme depicted in Fig.~\ref{fig1}, a pump pulse is used to prepare the electron in a coherent superposition of two Rydberg states of the hydrogen atom. These two states are represented by $|\chi_{n=9,l=8} \rangle$ and $|\chi_{n=10,l=9} \rangle$ with the angular quantum numbers $l=8$ and $l=9$, respectively. The superposition state has the form (atomic units are used throughout)
\begin{equation}\label{wf0}
|\psi(t) \rangle=\cos(\theta_R)|\chi_{n=9,l=8} \rangle+ \sin(\theta_R)e^{i\phi(t)}|\chi_{n=10,l=9}\rangle,
\end{equation}
where a relative phase information generated by a pump pulse is characterized by the phase $\phi(t)$. This phase is also a time-delay between the pump and the probe pulse, which is defined by $\tau=\phi/\Delta E$, where $\Delta E = E_{10} - E_9=$0.0012 a.u. is the energy difference between the two Rydberg states. Here, the energies are $E_{n=10}=$-0.005 a.u. and $E_{n=9}=$-0.0062 a.u.. The superposition state in Eq.~(\ref{wf0}) is determined by the polar angle $\theta_R$ and can be viewed in the Bloch sphere (see Fig. S1 in the Supplementary Information). After the pump pulse is turned off, a single-cycle pulse is introduced subsequently to coherently transport this mixture of states far away from the nucleus. The theoretical method used here is described in \textit{Methods}.

The spatiotemporal evolution of the electron density (integrated over the $\rho$-direction) encoding a phase information characterized by $\phi=0$ is shown in Fig.~\ref{fig1}(b). Note that the vector position of the electron is defined in cylindrical coordinates ($\rho,z$). The result is displayed for a peak intensity of 4.26 10$^{18}$ W/cm$^2$ and for an angular frequency of 0.057 a.u. (the corresponding duration is 1.3 fs). These parameters are chosen such that the final position of the transferred electron wavepacket is set at $z=\alpha(t_f)=1000$ a.u., where $\alpha(t)$ is the displacement vector illustrated in Fig.~\ref{fig1}(b) with with dashed blue line (see Eq.(\ref{disvec}) in \textit{Methods} for its analytical form). It is seen that the electron density follows a unidirectional path determined by the displacement vector. Moreover, the entire electron wavepacket is transferred far away from the origin to a well-defined region, in which a detector can be set for a readout process. In an experiment, the transferred electron can be captured and trapped inside a quantum dot for a sufficiently long time, such that it can be detected with conventional on-chip detectors~\cite{Bauerle2018}.

\begin{figure} [h!]
\centering
\includegraphics[width=16cm,height=10cm]{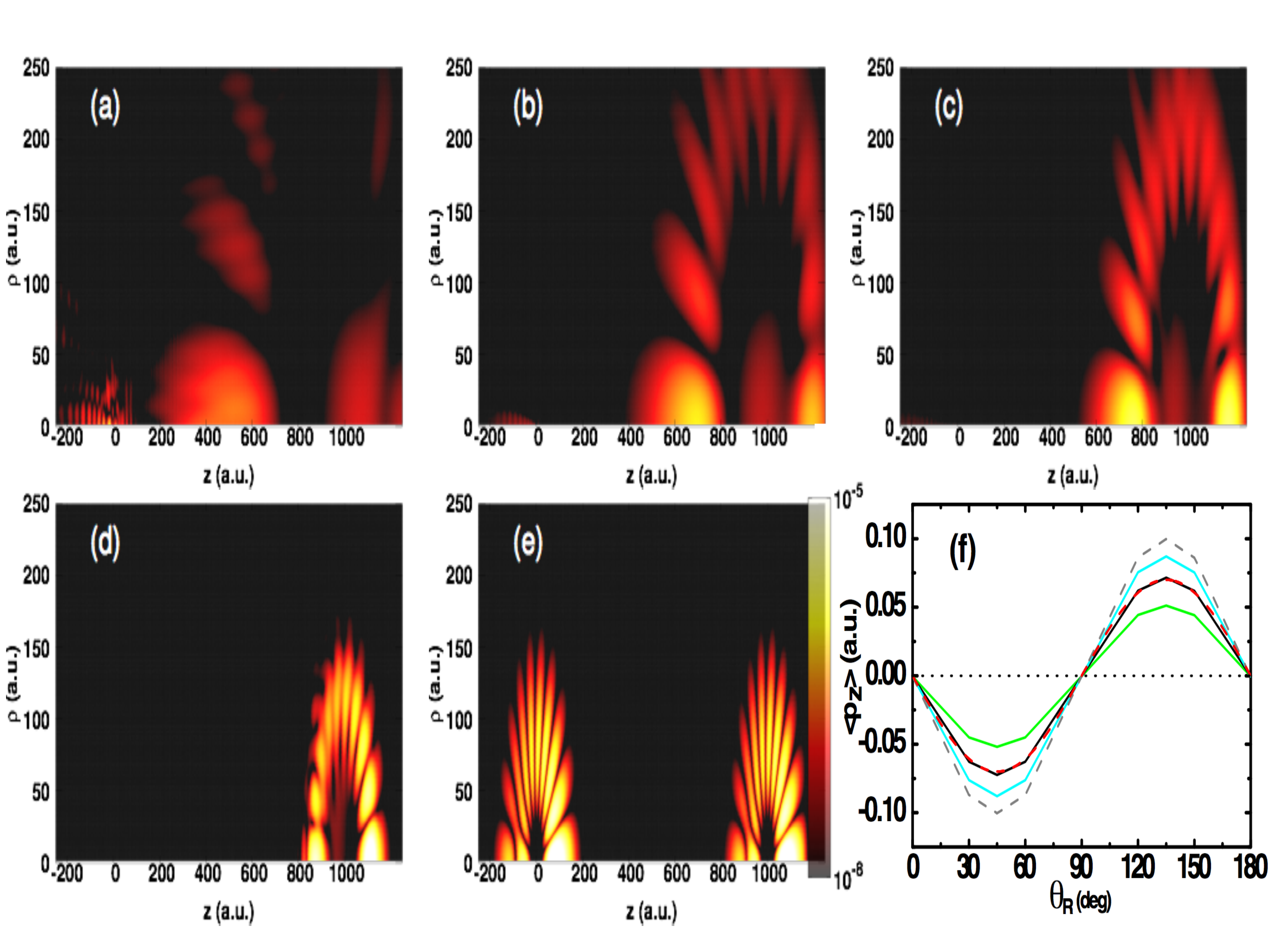}
\caption{\label{fig2} (a)-(e) Electron density in the $\rho$-$z$ plane at the end of the probe pulse for an initial phase of $\phi(t_i)=0$ and for the polar angle $\theta_R=\pi/4$, respectively at different angular frequencies and peak intensities: (a) $\omega$=0.0006 (4.78 10$^{10}$), (b) 0.00088 (2.45 10$^{11}$), (c) 0.00117 (7.66 10$^{11}$), (d) 0.0059 (4.78 10$^{14}$) and (e) 0.057 a.u. (4.26 10$^{18}$ W/cm$^2$). For reference, the initial state before the probe pulse is introduced is shown in the left-hand side of figure (e). (f) Expectation value of the momentum $<p_z>$ as a function of the polar angle $\theta_R$ at different initial phases $\phi=\pi/6$ (green solid line), $\phi=\pi/4$ (black solid line), $\phi=\pi/3$ (cyan solid line) and $\phi=\pi/2$ rad (dashed gray line). The peak intensity is fixed at 4.26 10$^{18}$ W/cm$^2$ and angular frequency is 0.057 a.u.. Also is shown the data from the simple model $b\cos(\phi(t_f) + \phi_0(t_f) )\sin(2\theta_R)$ for $\phi=\pi/4$ rad (dashed red line) (see text). The dotted line is for guiding.}
\end{figure}

The criterion here for observing a coherent transfer of the electron density without significantly altering its wavefunction is related to the frequency of the probing pulse which should be higher than that of the oscillation of the electron between the two Rydberg states. This is summarized in Fig.~\ref{fig2}, where changing the parameters of the single-cycle pulse (i.e., the peak intensity and angular frequency) results in a distortion of the electron wavepacket, as shown in Fig.~\ref{fig2} (a)-(c). Note that the change of the parameters is done such that the final position remains the same (i.e. $\alpha(t_f)=1000$ a.u.). In particular, it is seen in Fig.~\ref{fig2}(a) that at a smaller angular frequency of the pulse $\omega=\Delta E/2$=0.0006 a.u. (the corresponding peak intensity is 4.78 10$^{10}$ W/cm$^2$), the electron wavepacket is distorted and distributed in arbitrary directions with a small portion that remains localised at the origin. On the other hand, increasing the frequency beyond the value of $\Delta E$ results in a displacement of the entire electron wavepacket, although is distorted as displayed in Fig.~\ref{fig2}(b)-(c). Further increase of the angular frequency up to $\omega$=0.057 a.u. leads to the transfer of the electron wavepacket with no distortion, and in which the characteristics of the initial state are fully imprinted in the final state. This is shown in Fig.~\ref{fig2}(e) together with the electron density before the probing pulse is turned on. Initially, the density is localised at the origin and exhibits an asymmetry with respect to the $z$-axis. Clearly this asymmetry is the result of the coherent mixture between partial waves with the opposite parity (i.e., $l=8$ and $l=9$ waves). When the probing single-cycle pulse is introduced, the transported superposition state over a long distance preserves the signature of the coherent mixture. A similar result is found for a set of different phase information and different geometrical representation of the superposition state (see Figs. S2 and S3 in the Supplementary Information). On the contrary, in the case the criterion described above is not satisfied, the spreading of the electron wavepacket will make it difficult to access the initial phase information in the final state, as discussed below.

The observed spatial coherent dynamics of the superposition state can be exploited for retrieving the initial phase information. The starting point is to calculate the expectation value of the electron momentum $<p_z>$ at various polar coordinates $\theta_R$, which is equivalent to calculating the current density integrated over the electron coordinates. The results are shown in Fig.~\ref{fig2}(f) and are displayed after the single-cycle pulse is turned off for a set of different initial phases $\phi=\pi/6, \pi/4, \pi/3$ and $\pi/2$ rad (the corresponding time-delay is in the range of [10:32] fs). The result presented in Fig.~\ref{fig2}(f) can be seen as a time-resolved of the geometrical representation of the superposition state. In this figure, it is seen that the amplitude of the signal $<p_z>$ exhibits a coherent oscillatory behavior displayed in the polar angle $\theta_R$, which is found to be sensitive to the initial phase information $\phi$. This sensitivity indicates that the phase information can be mapped onto the $\theta_R$ coordinate, which can be detected in an experiment by measuring the electric current. 

Further details about the oscillatory behavior observed in Fig.~\ref{fig2}(f) can be seen in the picture of the mean momentum $<p_z>$ evaluated in the absence of the probing pulse, and is expressed as
\begin{equation}\label{pz}
<p_z> (\theta_R,\phi(t_i))= \rho_{ii}\cos^2(\theta_R) + \rho_{jj}\sin^2(\theta_R) 
                                            + \rho_{ij}\cos(\phi(t_i)) \sin(2\theta_R),
\end{equation}
where $\rho_{ij}=\langle \chi_{i}|p_z|\chi_{j} \rangle$ ($i$=$\{n=9,l=8\}$, $j$=$\{n=10,l=9\}$). Here the matrix elements $\rho_{ii}$ and $\rho_{jj}$ are zero. This is because of the symmetry of the density of the electron (not the wavefunction due to its parity) being in a single quantum state, and thus the integrand evaluated in momentum space $\rho_{kk}=\int p_z |\tilde{\chi}_{k}(\mathbf{p})|^2 d\mathbf{p}$ is odd, yielding zero for $\rho_{kk}$, $(k=(i,j))$. Here the wavefunction $\tilde{\chi}_{k}(\mathbf{p})$ is the Fourier transform of $\chi_{k}(\mathbf{r})$. Whereas the term $\rho_{ij}$ for $i \neq j$ is non-zero because of the asymmetry caused by the term $ \chi_{i}(\mathbf{r}) \; \chi_{j}(\mathbf{r})$, and which is imprinted in the  electron density as shown in Fig.~\ref{fig2}(e) and discussed above.

Now taken into consideration that the electron wavefunction after probing with the single-cycle pulse is preserved, as demonstrated above [cf. Fig.~\ref{fig2}] (see also Fig. S4 in the Supplementary Information), the induced $<p_z>$ can be expressed on the basis of Eq.~(\ref{pz}) and can be modified to take into account the effect of the probing field. This can be written in the simplified form
\begin{equation}\label{phase}
<p_z> (\theta_R,\phi(t_f)) = a+b\cos(\phi(t_f) + \phi_0(t_f) )\sin(2\theta_R).
\end{equation}
This formula, although is simple, it allows to retrieve the initial phase information as well as the action of the probing pulse. The expression in Eq.~(\ref{phase}) is obtained by assuming that the wavefunction $|\chi_{k} (t)\rangle$ in Eq.~(\ref{wf0}) is dressed and thus can be approximated by $|\chi_{k} (t)\rangle \; \rightarrow \; e^{-\Gamma/2} \; e^{-i\phi_k t} \; |\chi_{k} (t)\rangle$~\cite{Demekhin2013}, where $\Gamma$ is the decay and $\phi_k$ is the Stark shift of the state with the index $k$. Incorporating the dressed state $ |\chi_{k} (t)\rangle$ in Eq.~(\ref{pz}), leads to Eq.~(\ref{phase}), in which we introduce the parameters $b=e^{-\Gamma}$ and $\phi_0=\phi_i-\phi_j$. Here $a$ is a fitting parameter. The phase $\phi(t_f)$ is the retrieved phase information at the end of the probing pulse. By fitting the data in Fig.~\ref{fig2}(f) to the model in Eq.~(\ref{phase}), we are able to precisely retrieve the initial phase information. This is shown for an example of $\phi(t_i) = \pi/4$ rad, where the retrieved fitting parameters are $a=0$, $b=0.09$ and $\phi_0(t_f)=-0.033\pi$ rad. The data from this model is presented in Fig.~\ref{fig2}(f) with dashed red line, and is found to reproduce very well the data stemming from the TDSE. On the other hand, when the angular frequency of the single-cycle pulse is too small, which is the case displayed in Fig.~\ref{fig2}(a), the induced $<p_z>$ exhibits an oscillatory structure depending on the polar angle $\theta_R$ (see Fig. S5 in the Supplementary Information). But here the signal is asymmetric with respect to $\theta_R$, which makes it difficult to access the initial phase using the fitting symmetrical function in Eq.~(\ref{phase}). 
This demonstrates that coherent displacement of the electron wavepacket by means of a single-cycle pulse with a properly chosen frequency is a convenient strategy to learn precisely about the initial phase information. On the other hand, our results demonstrate that different initial phases have distinct current responses. These responses reflect the footprint of the coherent mixture of states, thus indicating that the quantum nature of the transferred information was preserved. Preserving quantum information is crucial in quantum communication and computation~\cite{Morton2008}. 

\begin{figure}[h!]
\centering
\includegraphics[width=16cm,height=6.5cm]{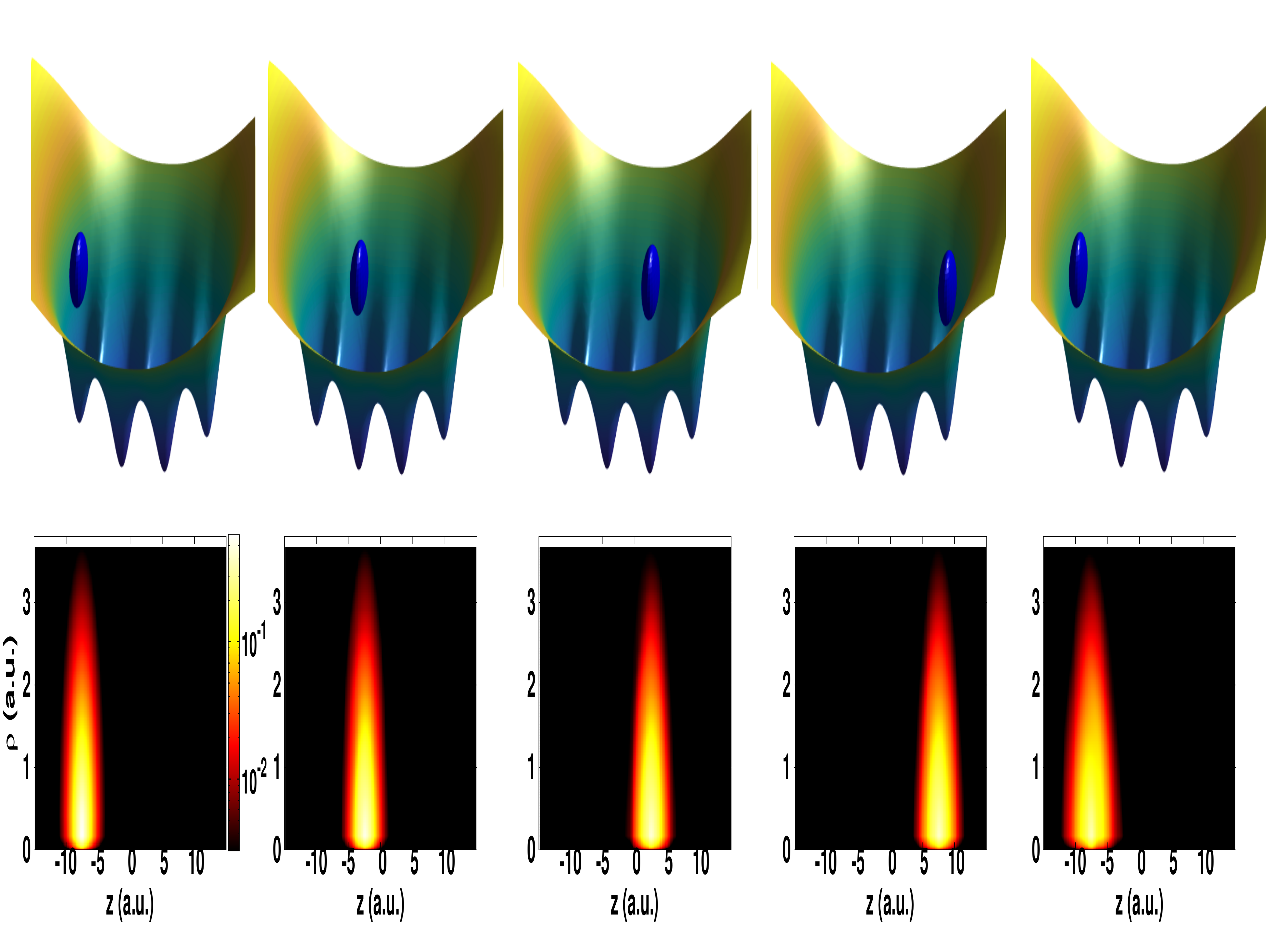}
\caption{\label{fig3} Illustration of a coherent transport of an electron wavepacket initially localised in a chain of four sites. Upper panel: Schematic diagrams showing the transport of the electron between neighbouring sites and back to the initial one. Lower panel: Electron density in the $\rho$-$z$ plane for the transported electron state using four subsequent single-cycle pulses. The peak intensity is fixed at 1.31 10$^{22}$ W/cm$^2$ and angular frequency is 6 a.u..}
\end{figure}

\subsection*{Electron transport in a chain of ions}
An extension of the mechanism to a chain of ions is shown in Fig.~\ref{fig3}. In this figure, we consider an example of a chain of four sites separated by a distance of 5 a.u.. The scenario is illustrated in the upper panel of Fig.~\ref{fig3} and enables us to demonstrate the possibility of the attosecond control of the directionality of a coherent moving of an electron back and forth in trapped ions. Here, the electron wavepacket is initially localised in the left site of the chain. We apply a sequence of attosecond single-cycle pulses of a peak intensity of 1.31 10$^{22}$ W/cm$^2$ and an angular frequency of 6 a.u. (total duration is 12.7 as). In this scenario the electron receives a momentum kick from the field during each single-cycle and as a consequence, it can coherently be transferred from one site to a neighbouring one. The results of observing the electron density in a specific site are shown in the lower panel of Fig.~\ref{fig3}. In the same figure (upper panel) is depicted a schematic illustration of the location of the electron in a given site. These results demonstrate the feasibility of of a coherent displacement of an electron wavepacket between neighbouring sites and even back to the initial site for an eventual readout again. The process is found to occur without significant change of the electron wavefunction. This is demonstrated by calculating the fidelity $F(t_i,t_f)=| \langle\psi(t_f)|\psi(t_i) \rangle|^2$, which here measures the \textit{closeness} of the state that is transferred back ($|\psi(t_f) \rangle$) and the original one ($|\psi(t_i) \rangle$); we find the fidelity to be 94 \%, which is very promising given the attosecond time-scale explored here. Note that the number of the displacement of the electron at this time-scale is limited by the width of the electron wavepacket in its initial state, which is here too small to enable several displacements with higher fidelity. To overcome this low fidelity issue it is desirable to use Rydberg atoms because of their long-life time. These obtained results, on the other hand, reveal the characteristic features of single-cycle pulses as a powerful tool for attosecond manipulation of quantum states, which is found to be achieved here with no significant spread out of the electron wavepacket.  

\subsection*{Entangled electron-electron state}
\begin{figure}[h!]
\centering
\includegraphics[width=12cm,height=6cm]{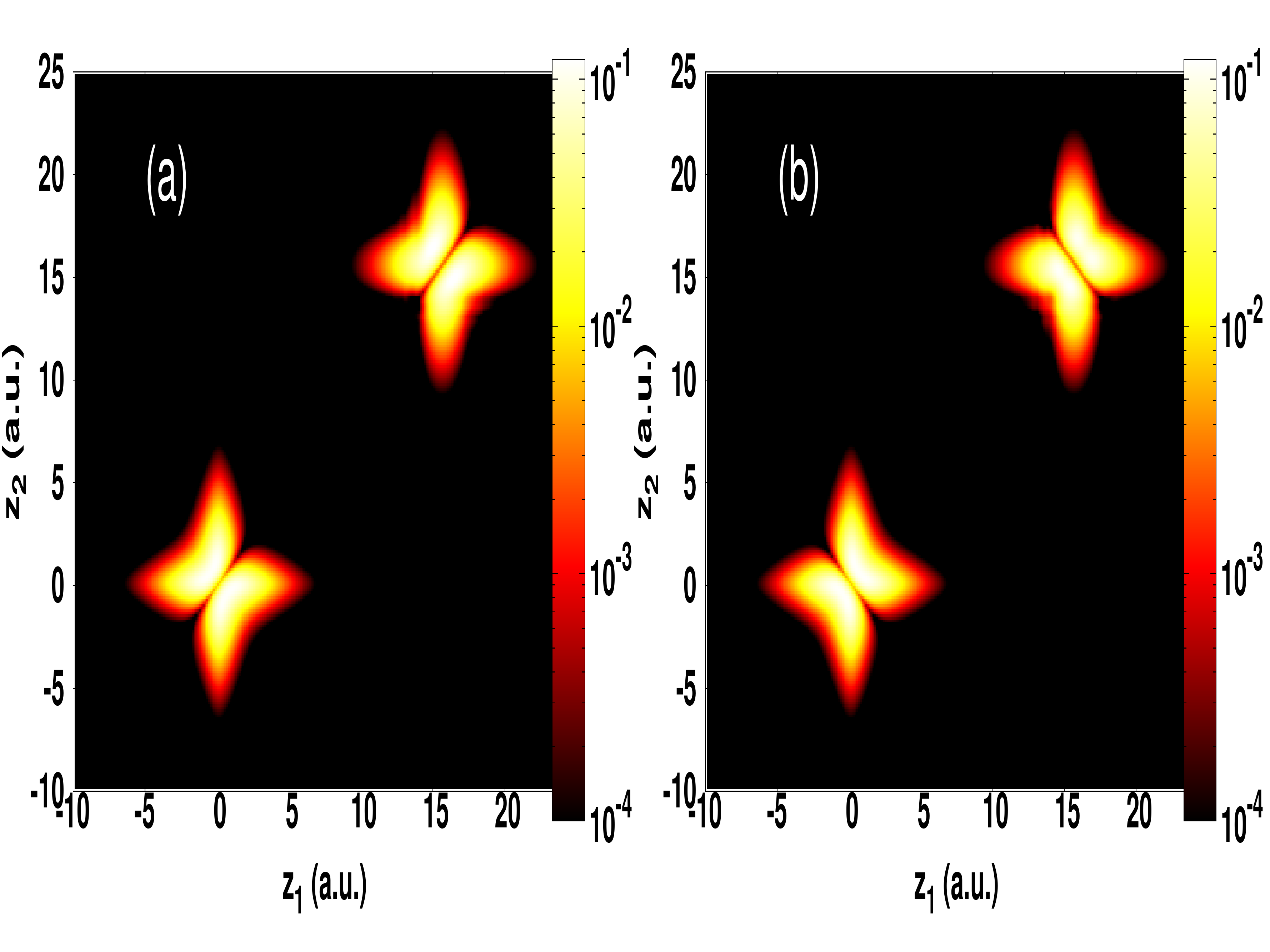}
\caption{\label{fig4}Electron density in z$_1$-z$_2$ plane at the end of the probing single-cycle pulse for calculations from a single spin state (a) and the triplet spin case (b) of helium atom. The peak intensity is 7 10$^{23}$ W/cm$^2$ and the angular frequency is 9.2 a.u.. For reference, the initial states before applying the probing pulse are shown at the origin.}
\end{figure}

The scenario demonstrated in this work is not limited to single-electron systems, but it can be extended to a two-electron spin dynamics. As an example, we show in Fig.~\ref{fig4} that such attosecond pulses can transport an entangled electron-electron state, in which an eventual information can be stored in the spin state. This is demonstrated for a prototype of helium atom using a reduced-dimensional model that takes into account the correlated electron-electron interaction as described in \textit{Methods}. Here, we consider a coherent displacement of two electrons initially prepared in a singlet spin state and the triplet case as well. This state is built-up according to
\begin{equation}\label{wf1s2s}
\psi(z_1,z_2)=  \frac{1}{\sqrt{2}}[ \chi_{1s}(z_1)\chi_{2s}(z_2) \mp \chi_{2s}(z_1)\chi_{1s}(z_2)],
\end{equation}
where the signs (+) and (-) correspond to the singlet spin state and triplet spin one, respectively. In Eq.~(\ref{wf1s2s}) the  one-electron eigenstates $\chi_{ns}(z)$ ($n=1,2$) are stemming from one-dimensional (1D) calculations using a soft potential model for helium ion as described in \textit{Methods}. We use a single-cycle pulse with a peak intensity of 7 10$^{23}$ W/cm$^2$ and an angular frequency of 9.2 a.u. (the corresponding duration is 8.3 as). At the end of the probing pulse, the result of observing the entangled electron-electron state far away from its origin is shown in Fig.~\ref{fig4} for both singlet spin state (cf. Fig.~\ref{fig4}(a)) and the triplet state case (cf. Fig.~\ref{fig4}(b)), and also for the ground state (see Fig. S6 in the Supplementary Information). For reference, the electron density of the initial state Eq.~(\ref{wf1s2s}) is also presented at the origin [cf. Fig.~\ref{fig4}(a) and (b)]. Once again, it is seen that the transport mechanism remains coherent and happens without altering the correlated electron wavepackets, in which the spin feature is preserved. Thus demonstrating the versatility of the use of these pulses. 

\section*{Discussion and Conclusions}

Here, we discuss the significance of probing with single-cycle pulses based on our findings. For instance, the observed unidirectional path followed by the electron wavepacket is an interesting feature, which could be exploited for determining the exact position of atoms inside materials, and that it is an ultimate form of atomic engineering~\cite{Su2019}. If experiments can verify these features it may lead to the emergence of new microscopes based on single-cycle pulses, such as the scanning transmission electron microscopy to characterize the precise atomic structure of materials (e.g. ref.~\cite{Su2019}).

Furthermore, the ability to manipulate coherently quantum states by means of single-cycle pulses offers possibilities for exploring quantum information and computation efficiently. This can be achieved by implementing the following pump-probe scheme. In this scheme, a pump laser pulse can be used to generate a bit of quantum information in an electron spin state, and an attosecond single-cycle pulse can be introduced to transfer this electronic state and store it in a nuclear spin state. Taken advantage of the characteristics of the nuclear spins in possessing a longer coherence time~\cite{Pla2013}, the quantum information can be then stored for some time before bring it back to the electron spin state for readout again. Moreover, by exploiting the properties of the electron-nuclear entanglement~\cite{Simmons2011}, it could be possible via the detection of the nuclear spins~\cite{Neumann2010,Morello2010} to readout with high-fidelity the initially stored quantum information. 

Last but not least, probing with single-cycle pulses can also be used for attosecond imaging of the electron wavepacket, thus allowing to reconstruct the full quantum dynamics. Imaging the electron wavepacket is motivated by recent experimental findings~\cite{Villeneuve2017}, where a full reconstruction of the continuum electron wavefunction was demonstrated. Most importantly, attosecond single-cycle pulses can also serve as a powerful tool for generating and characterizing attosecond electron pulses, which in turn can be exploited in connection with ultrafast electron microscopes and spectroscopes to image in time and space the atomic motion~\cite{Zewail2010}. Recently, there has been a great interest in generating ultrafast electron pulses. These pulses, which are produced so far either using electric field compression methods or relativistic electron sources, suffer from the charge space broadening which limits the temporal resolution~\cite{Manz2015}. Our work, therefore, adds new insights for the generation of attosecond electron pulses with the use of single-cycle pulses. 

With the state-of-the-art laser technology, it is possible to generate laser pulses with peak intensities, photon energies, pulse duration comparable to those used in the present work. For instance, the recent advent of x-ray free electron lasers (XFELs) allows to generate coherent x-ray radiation in the energy range 0.26-24.8 keV (wavelength 0.05-4.7 nm) with flux of about 5 10$^{33}$ photons per pulse with a duration less than 100 fs~\cite{XFEL}. This allows us to investigate the interactions between matter and high-frequency coherent light under extreme intensities of above 10$^{20}$ W/cm$^{2}$ (see e.g. Ref.~\cite{Yabashi2013}). Furthermore, the recently developed FEL facility and FERMI at Elettra can provide fully coherent ultrashort pulses with durations 10 to 100 fs~\cite{Eletrra}. In addition, the generation of isolated single-cycle pulses that operate at attosecond time scale becomes possible~\cite{Krauss2010,Jones2014,Sansone2006,Goulielmakis2008}. In this context, recent simulation has shown the possibility of generating attosecond single-cycle pulses with a duration of 50-100 as~\cite{Mak2018}. Therefore, the currently available super-intense coherent light pulses with high-frequency together with the emerging relativistic single-cycle pulses based technology~\cite{Nie2018}, make an experiment on attosecond coherent transport of an electron state feasible in the near future. 

At this point, our discussion presented here elucidates the broad impact and applicability of probing with single-cycle pulses. Such a versatile pulse is an emerging paradigm for advanced manipulation of the dynamical electron wavepacket. 

In conclusion, we have numerically demonstrated the potential of attosecond single-cycle pulses to provide unique insights into the coherent spatial electron dynamics. As an example, we have shown that the use of such pulses enables us to map out the spatiotemporal nature of the displaced electron wavepacket while it is oscillating coherently between two Rydberg states. In particular, it was found that the phase information encoded in the associated wavefunction can be mapped onto the polar coordinate in which the superposition state is defined. An extension to a singly charged ion in a string has shown the possibility of the attosecond control of the directionality of the coherent moving of the electron wavepacket between the neighbouring sites. The scenario is not limited to a single-electron system, but broadly applicable to many-body dynamics. Our study, therefore, provides a comprehensive picture of the role of a single-cycle pulse, as a versatile tool, in a variety of applications in modern electronics and ultrafast science, notably processing of quantum information and imaging at the attosecond timescale.

\section*{Methods}

The electron dynamics induced by a probing pulse is numerically simulated by solving the  the three-dimensional time-dependent Schr{\"{o}}dinger equation (3D-TDSE) in the velocity gauge, $ \Big[ H_0 + H_I(t) - i\frac{\partial}{\partial t}
\big]|\psi(t) \rangle=0, $ 
where $H_0=-\frac{\nabla^{2}}{2} +V(r)$ is the field-free Hamiltonian with the potential interaction $V(r)$, and $H_I(t)$ is the time-dependent interaction described within the dipole approximation and where the electric field is considered to be linearly polarized along the $z$-direction. Here, we use the same form of the vector potential as in~\cite{Micha2017b}; its derivative (i.e. the displacement vector) is depicted in Fig.~\ref{fig1}(b) with dashed blue line and has the form \cite{Micha2017b}
\begin{equation}\label{disvec}
\alpha(t) = [\frac{\sin(4\omega t)}{128} - \frac{\sin(2\omega t)}{32} +\frac{3}{32}\omega t]\frac{E_0}{4\omega^{2}},
\end{equation}
where $E_0$ is the maximum field strength and $\omega$ is the angular frequency. At the end of the probing pulse, the displacement vector is reduced to $\alpha(t_f)=1.178 U_p$, where $U_p=E_0/4\omega^{2}$ is the ponderomotive energy. According to this analytical formula, one can determine the strength of the single-cycle pulse required to displace the electron wavepacket to a desired position, while keeping the wavepacket undistorted by properly chosen the angular frequency. The TDSE is transformed into cylindrical coordinates and because of the cylindrical symmetry of the involved interactions, it is reduced to two coordinates ($\rho,z$). The equation is solved using a Split-operator method combined with a fast Fourier transform algorithm. Details about the method and numerical convergence have been discussed in our previous works~\cite{Agueny2018,Agueny2020}. 

In all the calculations performed in the present work, we use the dipole approximation to describe the interaction between the laser fields and the target. This approximation is mathematically justified when the spatial dimension of the electron wavepacket in its initial state is too small compared to the wavelength of the laser field, which allows to neglect the spatial variation of the fields. The approximation is known to break down in the X-ray regime at super-intense laser fields~\cite{Morten2018}, and also at the infrared regime~\cite{Bandrauk2015,Hartung2019} as has been observed and discussed recently. The breakdown manifests by a very small shift of high-energy electrons recorded in the photoelectron spectra~\cite{Morten2018}, and of slow electrons mapped in the momentum distribution. For instance, in Ref.~\cite{Bandrauk2015} the observed momentum shift due to the spatial dependence of the laser field was -0.009 a.u., which is too small to affect the general aspect of the dipole effect induced mechanisms. Taken into consideration the findings in previous works, we expect that the discussed displacement effect should hold even beyond dipole approximation. We have verified this statement by performing calculations, in which the spatial variation of the field is taken into account using a 2D-model similar to the ones used in Refs.~\cite{Bandrauk2015,Agueny2019}. The results are shown in Supplementary Fig. S7 and illustrate the dominance of dipole effects, thus confirming the validity of our findings.

For the transfer of a coherent superposition state, the calculations are performed on a grid of size $L_z=$4096 and $L_{\rho}=$512 a.u., respectively along the z-axis and $\rho$-axis, with the spacing grid $dz=d\rho=$0.25 a.u., (i.e., 16384 and 4096 grid points along z- and $\rho$-axis directions). The time step used in the simulation is $\delta$t$=$ 0.05 a.u. A smaller time step and spatial grid are used for the simulations in the case of a chain of ions (i.e., $\delta$t$=$0.001 a.u., $L_z=$512 $L_{\rho}=$256 a.u.). Here, the potential interaction between the electron and each site of the chain has the form
\begin{equation}\label{potChain}
V(z,\rho) = -\sum_{i=1}^4\frac{Z}{\sqrt{[z-(5-2i)R/2]^2 + \rho^2}},
\end{equation}
where $R=$5 a.u. is the inter-site distance and $Z=$0.8 is the nuclear charge. The ground state energy for the electron being localised at the left site is -0.322 a.u..

Calculations for the correlated electron-electron interaction are carried out in a reduced-dimensional model (i.e., one-dimension (1D) for each electron) with the use of a soft potential model describing the correlation interaction~\cite{Solve2011}
\begin{equation}\label{potHe1e}
V(z_1,z_2) = \frac{0.6317}{\sqrt{(z_1-z_2)^2+0.09168}},
\end{equation}
and the potential interaction~\cite{Solve2011}
\begin{equation}\label{potHe1e}
V(z) = -\frac{1.1225}{\sqrt{z^2+0.09169}}.
\end{equation}
With the use this model, the energies of the ground state and the first excited state of helium atom are well reproduced. The propagation of the initial state using the same form of the probing single-cycle pulse as described above is done by solving the 1D-TDSE for the correlated electrons. The numerical simulations are carried out in a symmetric spatial grid of size 512 a.u. having 2048 points.

The convergence is checked by performing additional calculations with twice the size of the box and a smaller time step. In our simulations, we did not incorporate an absorber, since we use a large box of simulation, which ensures that the entire electron wavepacket remains inside the box. We have however checked that using an absorber does not affect the results.

\section*{Data Availability Statement}
The datasets generated during the current study are available from the corresponding author on reasonable request.


\section*{Acknowledgements}
The author would like to thank Jan Petter Hansen for fruitful discussions and careful reading of the manuscript, as well as Ladislav Kocbach for helpful discussions. The computations were performed on resources provided by UNINETT Sigma2 - the National Infrastructure for High Performance Computing and Data Storage in Norway.

 
\section*{Additional Information}
{\bf Competing Interests:} The author declares no competing interests.

\end{document}